**Title**
- Second Harmonic Imaging Enhanced by Deep Learning Decipher

**Authors**

Weiru Fan,[1,†] Tianrun Chen,[1,†] Eddie Gil,[2] Shiyao Zhu,[1] Vladislav Yakovlev,[2,*] Da-Wei Wang,[1,3,*] Delong Zhang,[1,*]

**Affiliations**

[1]Interdisciplinary Center for Quantum Information and State Key Laboratory of Modern Optical Instrumentation, Zhejiang Province Key Laboratory of Quantum Technology and Device, and Department of Physics, Zhejiang University, Hangzhou 310027, China

[2]Department of Biomedical Engineering, Department of Physics and Astronomy, and Department of Electrical and Computer Engineering, Texas A&M University, College Station, TX USA 77843

[3]Zhejiang Laboratory, Hangzhou 311121, China

*Corresponding authors: yakovlev@tamu.edu; dwwang@zju.edu.cn; dlzhang@zju.edu.cn

[†]These authors contributed equally to this letter

**Abstract**

Wavefront sensing and reconstruction are widely used for adaptive optics, aberration correction, and high-resolution optical phase imaging. Traditionally, interference and/or microlens arrays are used to convert the optical phase into intensity variation. Direct imaging of distorted wavefront usually results in complicated phase retrieval with low contrast and low sensitivity. Here, a novel approach has been developed and experimentally demonstrated based on the phase-sensitive information encoded into second harmonic signals, which are intrinsically sensitive to wavefront modulations. By designing and implementing a deep neural network, we demonstrate the second harmonic imaging enhanced by deep learning decipher (SHIELD) for efficient and resilient phase retrieval. Inheriting the advantages of two-photon microscopy, SHIELD demonstrates single-shot, reference-free, and video-rate phase imaging with sensitivity better than $\frac{\lambda}{100}$ and high robustness against noises, facilitating numerous applications from biological imaging to wavefront sensing.

**MAIN TEXT**

**Introduction**

High-resolution optical imaging through transparent and scattering media strongly relies on accurate wavefront sensing and reconstruction [1-3]. Existing wavefront sensing approaches [4-6] are inherently based on interference and are sensitive to misalignment. Conventional optical phase microscopies also rely on interferometry that converts phase into measurable intensity variations, such as phase-contrast microscopy [7], digital holography [8-10], and diffraction phase imaging [11, 12]. The need of an additional reference beam perplexes the optical system and reduces its stability. Alternative approaches such as ptychography [13] or computational wavefront sensing [14], employ iterative computation which comes at the expense of multiple exposures. Real-world applications, such as biomedical imaging and remote sensing, however, require instantaneous wavefront reconstruction because of the highly dynamic imaging patterns.

In this report, we offer a nascent solution to this long-standing problem by utilizing the properties of the second-harmonic (SH) signal [15], which inherits the spatial phase of the incoming wavefront [16-18]. The wavefront change induced by the sample modifies the SH signal produced in a nonlinear crystal [18], which can then be used to retrieve the original phase image by solving the inverse problem of the second harmonic generation (SHG). A key advantage of using SH imaging contrast is the capability to encode abundant phase information from the fundamental beam, including the second derivative of phase. Such imaging retains details that are high in spatial frequency, providing the potential of artifact-free, high resolution and quantitative phase imaging capability. Furthermore, the proposed approach inherits multiple advantages of two-photon imaging techniques, including reduced photodamage and light-scattering, and high quantum efficiency when using near-infrared excitations.

A major challenge to this approach is the difficulty in solving the inverse SHG problem. The relation between the phase of the incident wavefront and the SH image can be solved by nonlinear Maxwell's equations. However, such interactions are rather complicated and can overwhelm the analytical methods (See Supplementary Materials). While numerical solutions based on the Fourier transform have been demonstrated, they are generally limited to simulations of simple, well-defined nonlinear interactions [19, 20]. For a more generalized nonlinear interaction, an approach capable of solving a specific relation without prior knowledge is needed. Notably, the recently developed deep learning approaches [21, 22], especially the convolutional neural network [23], provide an effective way to extract useful information without manual feature selections. By nature of the SHG process, the input and output have a complicated but definitive relation. With such advantages, we designed and implemented a new convolutional neural network to establish the relation between the SH intensity and the phase of the fundamental field. The new approach solves the inverse SHG problem and predicts the phase, which we term as second harmonic imaging enhanced by deep learning decipher (SHIELD).

SHIELD is advantageous for its capability of reference-free and lens-free phase imaging with high sensitivity. Unlike the conventional phase imaging or wavefront sensing that requires interference, SHIELD converts the phase image directly into SH intensity via SHG process (Fig. 1A). The reference-free setup removes the reference beam or the microlens array, enables a compact and highly efficient system by removing the possibilities of alignment and pointing errors, laser noises and instabilities. Furthermore, SHIELD signal is generated and captured directly from the changed wavefront, there is no need for a lens in the process, except when needed to increase SHG efficiency.

To extract the original phase distribution from the SH images, we developed a densely connected reconstruction network (DCRN), shown in Fig. 1B. Encode-decode structures [24] are widely-used algorithms for convolutional neural network. However, direct implementation of existing encode-decode structures [25] or simply stacking extra layers [26] have difficulty retrieving the original phase. It is because the SHG process encodes phase information in a nonlinear way, in which the features become hidden for common algorithms. To enlarge the feature space, we developed DCRN based on the densely connected layer structures [27] so that the hidden features can be extracted. As shown in Fig. 1B, each SH image goes through several composite layers, including convolution, Leaky ReLu, max pooling, and dense blocks, to generate high-dimensional feature maps up to 512 channels in order to successfully retrieve the original phase image (See Methods).

## Results

We implemented SHIELD experimentally and evaluated the performance of DCRN algorithm in terms of fidelity and robustness. Various phase images (Fig. 2A) were generated on a spatial light modulator (SLM) to shape the wavefront before the SHG crystal (See Methods). Although the acquired SH images are visually similar to a Gaussian profile, a quick singular value decomposition (Fig. 2A third column) showed clearly unique information in each feature map. During the training process, the parameters of DCRN were iteratively updated and optimized by comparing the prediction to the ground truth using Adam optimizer [28]. To quantitatively evaluate the prediction performance, we evaluated the fidelity of the prediction using Jaccard index [29] to quantify the similarity to the ground truth. A steady and consistent increase on accuracy in training and testing over the number steps was observed (Fig. 2B), indicating an optimization progress without under- or over-fitting. Importantly, we tested the sensitivity of the system by changing the depth of phase modulation (Fig. 2C). The resulting Jaccard Index fidelity remain high with modulation depth down to $\lambda/100$, i.e. 10.6 nm (see Supplementary Materials), demonstrating high sensitivity and scalability of the SHIELD system.

Robustness is a critical factor for deep learning-based imaging techniques since small noises sometimes leads to dramatical failures [30]. To address this issue, we tested our DCRN in three challenging scenarios (Fig. 3). First, we added white noises to the input SH images to disturb the DCRN prediction processes (Fig. 3A). High fidelity prediction was obtained despite the additional noises up to 30% of the average intensity of the SH images (quantified results in Supplementary Materials). Second, we address a common problem for image processing via deep learning, i.e., the tendency to make up an output. We input pure noises into the DCRN and found meaningless prediction results (Fig. 3B), demonstrating the reliability of DCRN. Third, we input artificial profiles with 1, 2, and 3 Gaussian centers that are similar to real SH images (Fig. 3C). DCRN output was not fooled to register images with any meaningful characters. Further tests on partial over-exposure inputs were demonstrated in Supplementary Materials. In a word, the robustness of DCRN is sufficiently high to discard unrelated information, including pixel noise and artificial input mimicking SH images. Furthermore, we tested the prediction performance to be similar on different image datasets when trained by either EMNIST and Quick Draw Dataset (Fig. 3D), showing generalization ability on various datasets (Supplementary Materials).

SHIELD also demonstrates high speed imaging and processing due to the compact and efficiency-optimized network structure. On average, it took 35 seconds for SHIELD to finish retrieving 800 images with 512×512 pixels using a personal computer equipped with a consumer graphics card (see Methods). It is equivalent to 44 ms per image, i.e. 22.8 frames per second, near video-rate.

## Discussion

It is worth discussing the potential capabilities of SHIELD as an emerging platform for wavefront sensing and phase imaging. While SHIELD relies on nonlinear optical interactions, we have demonstrated that a reduction of SH signal by almost three orders of magnitude has no significant effects on the outcome (see Supplementary Materials). To further enhance the sensitivity, more efficient nonlinear optical crystals can be used [31], as well as more sensitive detectors, such as single-photon sensitive sCMOS.

We note that other parameters, such as incident angle and polarization, could also affect our imaging capabilities. By tilting the nonlinear crystal or changing the ellipticity of polarization, we found little impact on the performance (quantified results in Supplementary Materials). In more general case of an arbitrary polarization, a polarizer can be used to split the incoming beam into two independent channels for imaging in two orthogonal polarizations. Furthermore, given the high sensitivity and dynamic range of SHIELD, it can also detect subtle changes in phase. Previously, we reported a bond-selective transient phase imaging by introducing infrared pulses and detecting the subsequent phase changes caused by absorption with sub-µs temporal resolution [32]. With its single shot capability and fs laser pulse detection, SHIELD is expected to enable chemically selective phase imaging with femtosecond temporal resolution, which is a 6 order of magnitude improvement. Another advantage of SHIELD for biological imaging is that it reduces phototoxicity by probing with lower energy photons.

Recently, with the capability of deep learning, a lens-free imaging approach has been demonstrated [33], in which the phase information is encoded in the interference-based diffraction patterns that evolves along the propagation direction. Thus, it is difficult to provide consistent patterns, causing the system susceptible to the distance to the sensor. In contrast, SHIELD approach is fundamentally different, in which the phase information is encoded in the SHG signal in a definitive and consistent manner, providing high stability and robustness against imaging parameters such as distance to the camera, or beam divergence. Furthermore, convolutional neural network based phase microscopy has been demonstrated [34], showing extended depth-of-field and phase recovery capability. SHIELD as a novel nonlinear optics based technique, offers unique advantages over the existing deep learning strategies.

In conclusion, we demonstrated a new phase imaging paradigm utilizing SH intensity imaging and deep learning. High performance, high robustness phase reconstruction has been demonstrated using our proposed DCRN structure, which provides a promising platform in the realm of biomedical imaging, medical diagnoses, materials sciences, and optical communications.

## Materials and Methods
### Experimental setup
An infrared pulsed laser was used for the experiments (1064nm, 330fs; pump: SPIRIT 1040-16_30-HE; OPA: Orpheus-HP, Spectra-Physics). The power of the laser was controlled by a half waveplate and polarizing beam splitter. The laser passes through a 1064-nm bandpass filter of 10mm bandwidth, and through a laser collimation system formed by a pair of lenses. The laser spot was shaped by an iris. An SLM (X13139-09, Hamamatsu) was used to modulate the phase of the laser. The modulated light was then collected and focused into a nonlinear crystal (periodically poled potassium titanyl phosphate, Raicol), in which the SHG signal was generated. The SHG signal was filtered by a 532-nm bandpass filter, and the intensity information was recorded by a CCD (Prosilica GT1910, AVT). In the experiment, the average power of the fundamental field before the crystal was 1.4 mW/cm$^2$. The average power of the generated SHG signal was 9.3 µW/cm$^2$ without phase modulation. An image of $512 \times 512$ pixels and 8-bit grayscale intensity distribution was recorded by the CCD. The modulation region was $6 \times 6$ mm squared area, which corresponded to the laser spot. The crystal was working at a temperature of 30 degrees Celsius. The image acquisition interval of each frame was 0.5 s (See Supplementary Materials).
### DCRN implementation
Structure: Upon the input of $512 \times 512$ images, a compression feature extraction image of $256 \times 256$ of 3 channels was generated by a $2 \times 2$ convolution without activation function, then transformed through network layers including $3 \times 3$ convolution kernels, Leaky ReLu activation functions, dense block units, and max-pooling to form deeper feature maps, defined as encoding. The feature maps then went through a decoding process, where layers were concatenated with the feature maps of the same size in the encoding process ($32 \times 32 \times 256$, $64 \times 64 \times 128$, $128 \times 128 \times 64$, and $256 \times 256 \times 32$ blocks, respectively), operated by $2 \times 2$ deconvolution, $3 \times 3$ convolution, Leaky ReLu activation function, and dense block unit (Fig. 1b). The feature maps were then transformed into an 8-channel image by a $1 \times 1$ convolution, then into a $512 \times 512$ image of 2 channels by a tensor operation of depth to space before the final prediction image. In the process, feature maps with deep channels interleaved with operations, i.e., dense blocks, were used, with operations including $1 \times 1$ convolution prior to $3 \times 3$ convolution. 5-layer and 4-layer dense blocks were used for down- and up-sampling.

Training: After the network is built, the training and evaluation process was performed under the Tensorflow framework on a server (10 GB RAM, Xeon E5, 2080 Ti, CUDA 10.0). Jaccard index was used to evaluate the similarity between the predicted image and the original phase image, and cross-entropy as the loss function. Handwritten characters from the Extended Modified National Institute of Standards and Technology (EMNIST) dataset and Quick Draw Dataset were used as the input to the SLM and the ground truth image for DCRN. For each dataset collected, 90% and 10% were used for training and testing, respectively. Each set was fed into the DCRN and trained with 50 epochs by the Adam optimizer with dynamic learning rates. The DCRN was trained by optimizing the parameters that minimize the loss function. More details are provided in the Supplementary Materials.

**Acknowledgments**: **Funding:** This research was in part supported by the National Science Foundation of China (11934011, 11874322), the National Key Research and Development Program of China (2019YFA0308100, 2018YFA0307200), the Zhejiang Province Key Research and Development Program (2020C01019), the Basic Research Funding of Zhejiang University, the Information Technology Center of Zhejiang University, Fundamental Research Funds for the Central Universities of China, the Major Scientific Research Project of Zhejiang Lab (No.2019MB0AD01，20190057), the National Science Foundation (DBI-1455671, ECCS-1509268, CMMI-1826078), the Air Force Office of Scientific Research (FA9550-15-1-0517, FA9550-20-1-0366, FA9550-20-1-0367), the National Institutes of Health (1R01GM127696-01), and the Cancer Prevention and Research Institute of Texas (RP180588). **Author contributions:** W.F and D.W.W. conceived the idea. W.F., T.C., D.Z., D.W.W. designed the project. W.F. performed the experiments. T.C. designed the algorithm. W.F., T.C., E. G., D.Z., D.W.W. and V.V.Y. analyzed the data. W.F, T.C., D.Z. and D.W.W. wrote the manuscript with input from E. G., V.V.Y. and S. Z. All authors discussed the results and commented on the manuscript. **Competing interests:** The authors claim no conflict of interests.

**Figures and Tables**

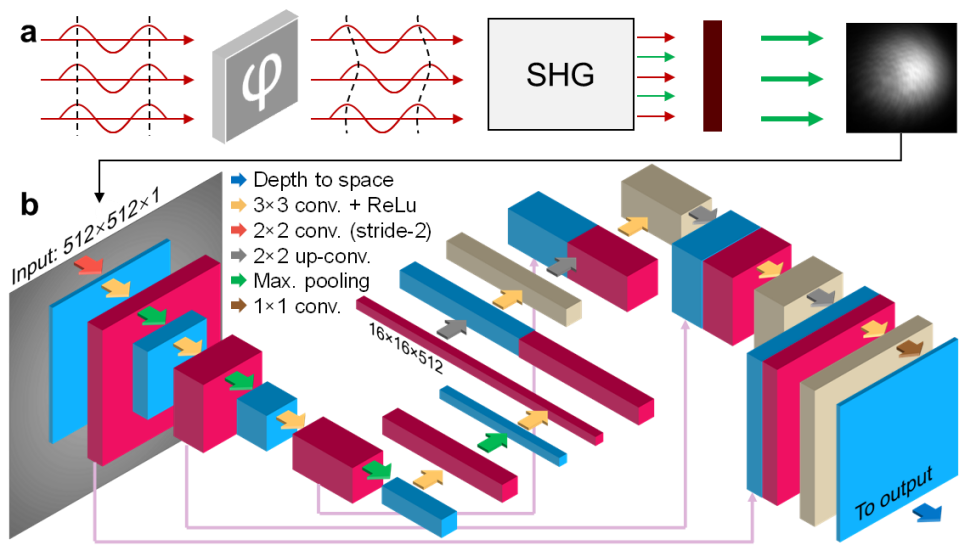

**Fig. 1. SHIELD scheme illustration.** (**A**) Near-infrared pulses containing the phase of a sample are sent through an SHG crystal. The frequency-doubled light is recorded by a camera with the fundamental light filtered out. (**B**) Densely connected reconstruction network. Each 3D block represents a multichannel feature map. Each arrow indicates an operation as shown in the legend. The final output image is omitted for visual clarity.

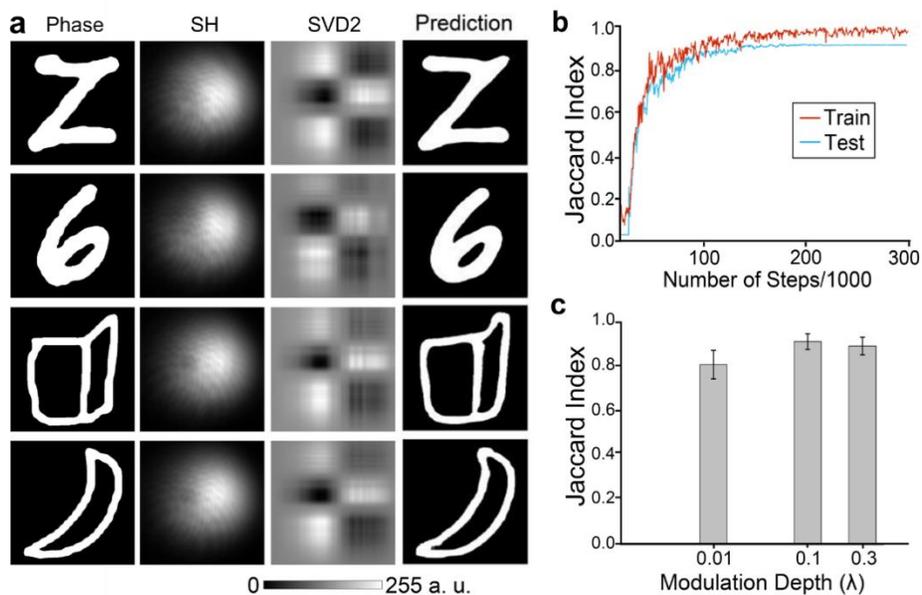

**Fig. 2. Performance characterization of SHIELD.** (**A**) Randomly selected ground truth phase images (first column), the corresponding raw SH images (second column), singular value decomposition feature maps (third column), and the corresponding prediction results (fourth column). (**B**) Training and testing performance over steps in terms of Jaccard index. (**C**) Performance evaluation of the test dataset for various phase modulation depths down to 0.01 λ. Error Bar: ± standard deviation.

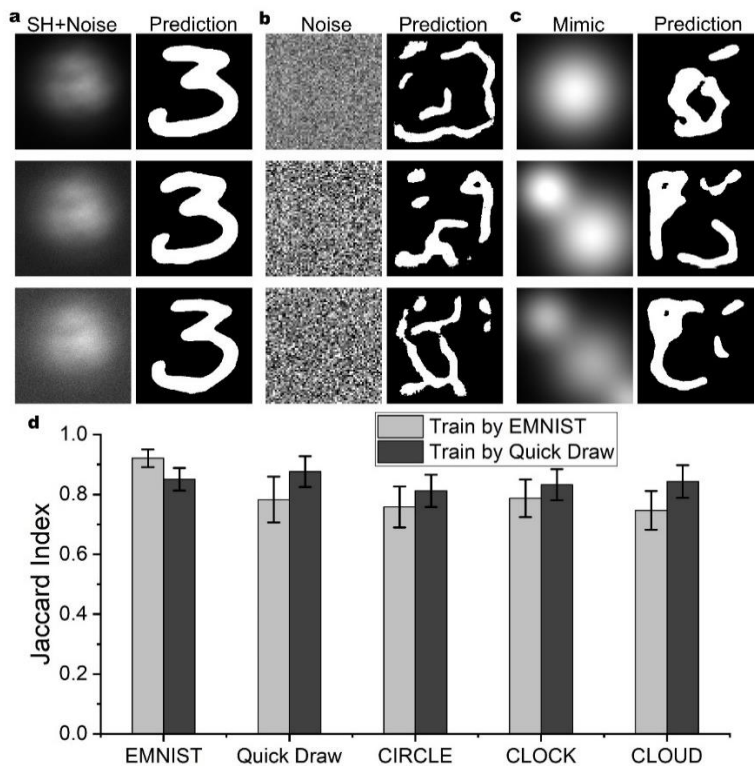

**Fig. 3. Robustness against noises, artificial inputs, and training sets.** (**A**), Input SH images with Gaussian noises added (left column) and the prediction results (right column). (**B**), Pure noises as input (left column) and the prediction results (right column). Noise images were downsampled by 8 for visual clarity. (**C**), Artificial Gaussian images as input (left column) and the prediction results (right column). (**D**) Performance on various datasets when trained by EMNIST (gray) or Quick Draw Dataset (dark gray).

# Supplementary Materials

## 1. Theoretical model of SHG

SHG in homogenous medium can be described by three-dimensional wave equation [15]

$$\nabla^2 \tilde{E}_j(\vec{r},t) - \frac{n_j^2}{c^2}\frac{\partial^2 \tilde{E}_j(\vec{r},t)}{\partial t^2} = \frac{1}{\varepsilon_0 c^2}\frac{\partial^2 \tilde{P}_j(\vec{r},t)}{\partial t^2} \tag{S1}$$

where $j = 1$ and 2, represent fundamental and SHG field, respectively. $\epsilon_0$ is the permittivity in free space, and $n$ is the refractive index. $c$ represents the speed of light in a vacuum. We use the complex amplitude and phase to represent the electric field

$$\tilde{E}_j(\vec{r},t) = \tilde{A}_j(\vec{r})\exp[i(k_j z - \omega_j t)] \tag{S2}$$

and the nonlinear polarization

$$\tilde{P}_j(\vec{r},t) = \tilde{P}_j(\vec{r})\exp(-i\omega_j t) \tag{S3}$$

The envelope of a forward-travelling wave pulse is assumed to be varying slowly in time and space compared to a period or wavelength, i.e. the complex wave amplitudes $\tilde{A}_j(\vec{r})$ may be assumed slowly varying in the z direction, i.e. $|\partial^2 \tilde{A}_j / \partial^2 z| \ll |k_j \partial \tilde{A}_j / \partial z|$. By inserting Eqs. S2, S3 into the Eq. S1, we obtain a SHG equation

$$\frac{\partial \tilde{A}_j(\vec{r})}{\partial z} = \frac{i}{2k_j}\nabla_\perp^2 \tilde{A}_j(\vec{r}) + \frac{ik_j}{2\varepsilon_0}\tilde{P}_j(\vec{r})\exp(-ik_j z) \tag{S4}$$

We decompose the complex amplitude into amplitude and phase, and consider the nonlinear polarization of the SHG process [15,18]

$$\tilde{A}_j(\vec{r}) = A_j(\vec{r})\exp[i\varphi_j(\vec{r})] \tag{S5}$$

$$\tilde{P}_1(\vec{r}) = 4\varepsilon_0 d_{eff}\tilde{A}_1^* \tilde{A}_2 \exp[i(k_2 - k_1)z] \tag{S6}$$

$$\tilde{P}_2(\vec{r}) = 2\varepsilon_0 d_{eff}\tilde{A}_1^2 \exp(2ik_1 z) \tag{S7}$$

Each nonlinear crystal is designed for a very specific orientation of input and output polarizations, so that the $d_{eff}$ is specific to each type. By substituting Eqs. S5-S7 into Eq. S4 and separating the real and imaginary parts, leads to the following system of equations describing the SHG process:

$$\partial_z A_j(\vec{r}) = -\frac{1}{2k_j}\{2[(\partial_x A_j)(\partial_x \varphi_j) + (\partial_y A_j)(\partial_y \varphi_j)] + A_j \nabla_\perp^2 \varphi_j\} + (-1)^j F_j \sin(\Delta k z + \Delta\varphi) \tag{S8}$$

$$A_j \partial_z \varphi_j(\vec{r}) = -\frac{1}{2k_j}\{A_j[(\partial_x \varphi_j)^2 + (\partial_y \varphi_j)^2] - \nabla_\perp^2 A_j\} + F_j \cos(\Delta k z + \Delta\varphi) \tag{S9}$$

where $\Delta k = k_2 - 2k_1$, $\Delta\varphi = \varphi_2 - 2\varphi_1$, and $F_1 = 2k_1 d_{eff} A_2 A_1$, $F_2 = k_2 d_{eff} A_1^2$. $A$ and $\varphi$ represent amplitude and phase, respectively. $d_{eff}$ is the effective nonlinear coefficient. Note the second derivative of phase in Eqs. S8 and S9.

Even in such simplified form, it is difficult to solve the nonlinear equations for an arbitrary input phase distribution, because the key variable, phase gradient, is often affected by the 2-π phase ambiguity making it a challenging problem.

## 2. Relation between physical processes and densely connected reconstruction network (DCRN)

In the previous section, we have shown that the equation governing the SHG process is computationally challenging, and no simple analytical solution exists. Therefore, we consider using numerical method to solve this problem. Specifically, we use a common optimization scheme, which considers the following equation

$$p = \arg\min\left\{\left\|A_{out} - |T \times E_{in}|\right\|^2 + \delta\right\} \tag{S10}$$

where $E_{in}$ is complex electric field of fundamental light, and $A_{out}$ is the amplitude of the SHG field. $T$ represents the transmission matrix for the SHG process, and $\delta$ is regularization term, in which prior knowledge is introduced to eliminate the influence of noises in the experiment. $p$ is an objective function.

For DCRN, the goal is to recover the phase information of the fundamental field. Therefore, we need to consider the reverse process in Eq. S10. In other words, the intensity distribution of the SH should be the input of DCRN, and the output of DCRN should be the predicted phase distribution of fundamental field. Mathematically, this process can be expressed as the following equation

$$\hat{p} = \arg\min \sum_{n=1}^{N} \left\| g_n - \prod_\theta H_\theta \times f_n \right\|^2 + \Delta \tag{S11}$$

where $\hat{p}$ represents the objective function used to represent the performance of network parameters. $f_n$ and $g_n$ represent the input and output of DCRN, respectively. $\Delta$ is regularization term of DCRN on the parameters with the aim of avoiding overfitting. $H$ represents every mathematical operation and transformation in the network layer, and $\theta$ represents the serial number of operations. The structure of our network is represented by the set of parameters $H$ in Eq. S11. $H$ can be expressed as convolution, pooling and other related operation units. For DCRN, the filter in the traditional algorithm is obtained through learning through a large number of training samples.

## 3. Details of DCRN implementation

The input of DCRN is the intensity image captured by the CCD, and the expected output is the reconstruction of phase distribution. In DCRN, each input image goes through an encode path (Fig. S1(a) upper left). The encode path starts from down-sampling by a convolution layer with no activation function that reduces the resolution of the input image by half, allowing the network selectively retain useful information for reconstruction phase information in subsequent layers. This approach effectively reduces the memory consumptions and computational complexity required in successive layers. The input image then goes through several layers of convolution, Leaky ReLu, max pooling, and dense blocks, generating feature maps with more channels but reduced resolution. The resultant high dimensional low-resolution feature maps then go through a decode path (Fig. S1 (a) upper right), which consists of the concatenation of feature maps from corresponding layers at encode path, followed by a series of up convolution, convolution and dense block. At the end of the decode path are two feature maps corresponding to the output phase image and its counterpart.

By visualizing the output of each layer, we can better understand the operations performed at each layer of DCRN (Fig. S1 (b)). The input image first goes through a "encoding" process for feature extraction, which transforms the image through continuous convolution and pooling operations. In this process, DCRN effectively learns local and global features useful for reconstruction of phase image. In the other "decoding" process, the low-resolution feature map of high dimension is fused with features of different dimensions in the "encoding" process and deconvolution is performed, and each layer keeps getting closer to the desired target image. We show the similarity between the average value of the feature map in each primary layer and the original target image (Fig. S1 (c)).

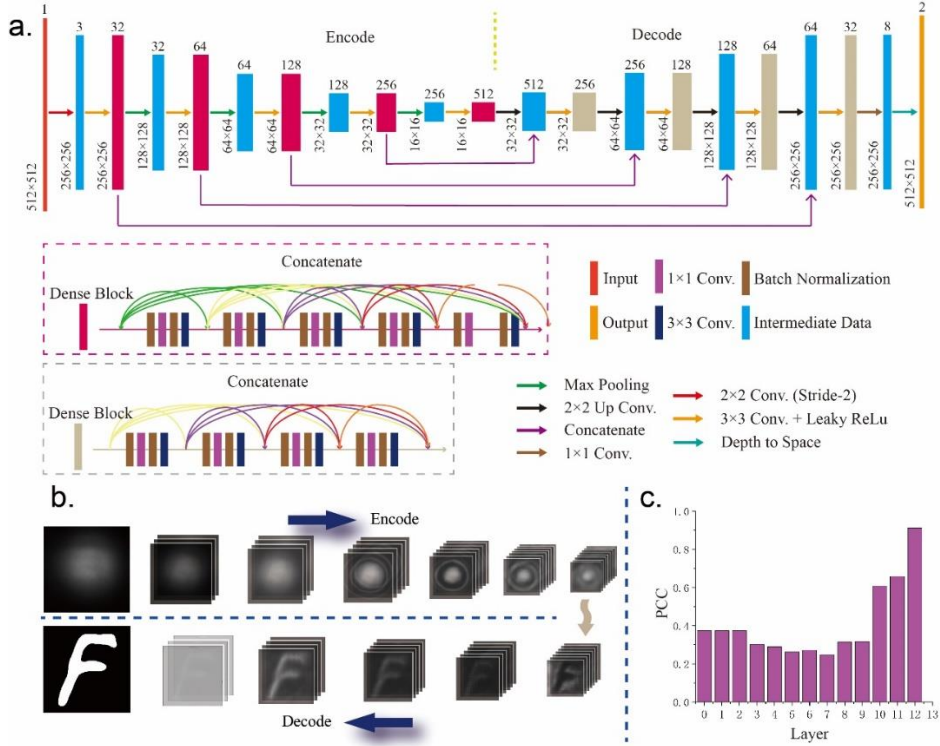

**Fig. S1. Detailed structures of SHIELD**. (a) Structure of DCRN. Each box corresponds to a multi-channel feature map. The number of channels is shown at the top of the box. The dotted boxes represent the structure inside the dense block. The encode and decode stages are shown by different dense block structures. (b) The intermediate feature maps in encode and decode process. (c) The curve quantitatively describes the similarity between activation maps of each layer and the original phase image using Pearson correlation coefficient (PCC). The trend of PCC is to decrease in the encoding process and increase in the decoding process.

During the training process, the parameters of DCRN are adjusted constantly. Batch size is set to 1, which means after each sample input into the network, the parameters are updated accordingly (a "step" in Fig 2(b)). This approach can effectively use the information of each sample to promote learning efficiency, while avoiding the excessive influence of bad samples on the training process and allowing high-resolution images input under limited physical graphic memory. Adam optimizer is used in the process of updating the network parameters [28]. By dynamically adjusting the learning rate, the Adam optimization method allows rapid convergence and capability to handle big dataset in our tasks. Table S1 shows the size and number of network parameters.

**Tab. S1. List of parameters for the proposed DCRN.**

| Parameters of Neural Network |
| --- |

| Layer | Type | Shape | Kernel Parameters Number | Bias Parameters Number | BN Parameters Number |
|---|---|---|---|---|---|
| conv0 | float32 | 2x2x3x3 | 36 | 3 | |
| conv1 | float32 | 3x3x3x32 | 864 | 32 | |
| DB1 | float32 | | 76800 | 384 | 864 |
| conv2 | float32 | 3x3x32x64 | 18432 | 64 | |
| DB2 | float32 | | 307200 | 768 | 1728 |
| conv3 | float32 | 3x3x64x128 | 73728 | 128 | |
| DB3 | float32 | | 1228800 | 1536 | 3456 |
| conv4 | float32 | 3x3x128x256 | 294912 | 256 | |
| DB4 | float32 | | 4915200 | 3072 | 6912 |
| conv5 | float32 | 3x3x256x512 | 1179648 | 512 | |
| DB5 | float32 | | 19660800 | 6144 | 13824 |
| Upconv1 | float32 | 2x2x256x512 | 524288 | | |
| conv6 | float32 | 3x3x512x256 | 1179648 | 256 | |
| DB6 | float32 | | 3932160 | 2560 | 5120 |
| Upconv2 | float32 | 2x2x128x256 | 131072 | | |
| conv7 | float32 | 3x3x256x128 | 294912 | 128 | |
| DB7 | float32 | | 983040 | 1280 | 2560 |
| Upconv3 | float32 | 2x2x64x128 | 32768 | | |
| conv8 | float32 | 3x3x128x64 | 73728 | 64 | |
| DB8 | float32 | | 245760 | 640 | 1280 |
| Upconv4 | float32 | 2x2x32x64 | 8192 | | |
| conv9 | float32 | 3x3x64x32 | 18432 | 32 | |
| DB9 | float32 | | 61440 | 320 | 640 |
| conv10 | float32 | 1x1x32x8 | 256 | 8 | |
| Total Trainable Variable | | | 35296687 | | |

The first column indicates the layer architecture of the neural network. "conv0" represents the first convolution layer, "DB1" represents the first Dense Block. All parameters are float32 type. The third, fourth, and fifth column indicates the parameter numbers of the Kernel, Bias and Batch Normalization, respectively. A total number of 35,296,687 variables are trained in the neural network.

**4. Additional results and analysis**

To demonstrate the sensitivity of SHIELD, we tested the system using phase targets with modulation depth of 1/100 wavelength, selected images are shown in Fig. S2.

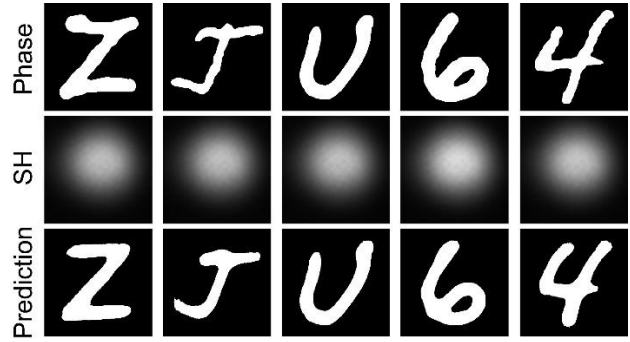

**Fig. S2. Prediction results for modulation depth of 0.01 λ.**

We tested the DCRN's performance by introducing additional Gaussian noise to the input image, calculated by σ/mean, where σ is the standard deviation of Gaussian distribution and mean equals to the averaged intensity for every pixel in the image before noise was added. The reconstruction (Fig. S3) showed good Jaccard index even at the noise level of 30%.

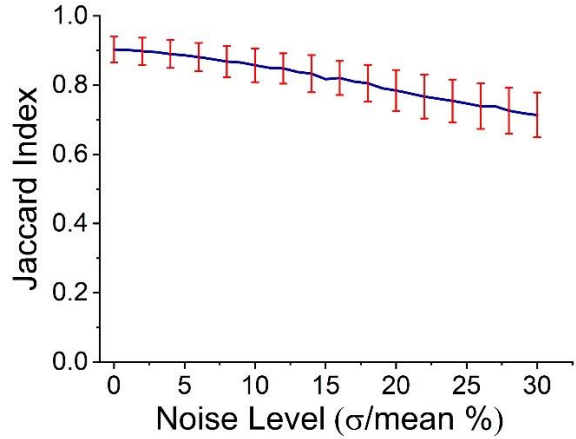

**Fig. S3. Evaluation results for images with different level of noise added.**

The robustness that SHIELD possess enables various applications. SHIELD can be applied when there is a small misalignment of the incident beam and the crystal, as long as the SHG signal can be produced. Such the property is shown in Fig. S4 (a), where we tilt the incident beam for ~1° both horizontally and vertically before it passes through the crystal, and compared the reconstruction result with the direct-incident scenario. We also tested the system under various laser power, which affects both the number of photons that CCD can receive and the conversion efficiency of SHG crystal. The result is shown in Fig. S4(b), where the reconstruction accuracy remains high under low power input, indicating the capability of SHIELD working under low laser power, which is especially important in biomedical imaging where a probe light with low power penetrate the sample would reduce the photobleaching/ phototoxicity.

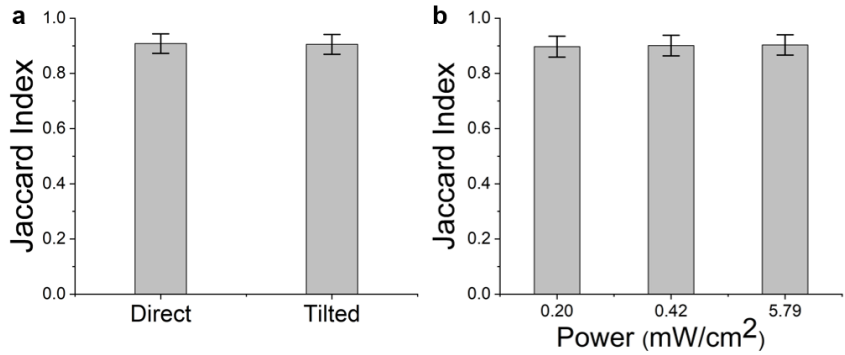

**Fig. S4. Additional results of SHIELD with adversely changed conditions**. (a) Test results for tilted incident beam. The reconstruction accuracy quantified by Jaccard Index is the same for direct incident and light being tilted (p<0.01), indicating careful alignment is not essential in SHIELD. (b) Test results for different power of the incident beam of crystal. By using the different neutral density filter before the camera, the average intensity and the exposure time remains constant with lasers of different power level. The reconstruction accuracy quantified by Jaccard Index is not affected by the laser power. (p<0.01)

Furthermore, when light passes through a sample, its polarization might alter. To emulate this phenomenon, we tested SHIELD with linearly polarized light 45° illuminated the SLM, and the modulated light is elliptically polarized. The results in Fig. S5 shows that high accuracy reconstruction can be made with different polarization.

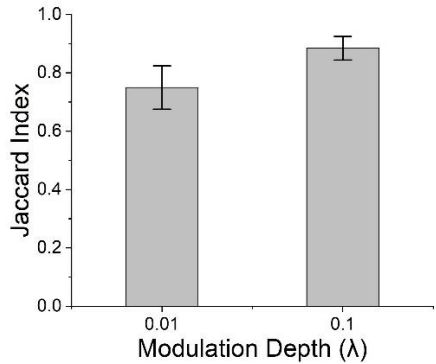

**Fig. S5**. **Test result for difference polarized incident light modulated by SLM**. Different modulation depth corresponds to different elliptic polarization.

Moreover, we tested the DCRN with images that have partially lost information due to, in our case, overexposure. A longer exposure time of CCD will reduce the inherent fluctuation noise and capture more details of the desired signal. Therefore, we set our CCD to work at the longest possible the exposure time (15ms) without pixel saturation. However, according to the Eq. S8, S9, different phase modulation of the fundamental field will result in change in conversion efficiency of the SHG process. If a specific phase distribution of fundamental field that effectively improves the conversion efficiency go through the SHG process, the output light might be too intense, causing images captured partially overexposed. In this case, the effective information of the image will be partially lost. For neural networks, such samples put an extra hindrance in learning and recovering phase, while our proposed DCRN still perform well with such inputs. Images with oversaturated areas can still work for both training and generating predictions. Some prediction results of the overexposure samples are shown in Fig. S6.

Page **15** of **17**

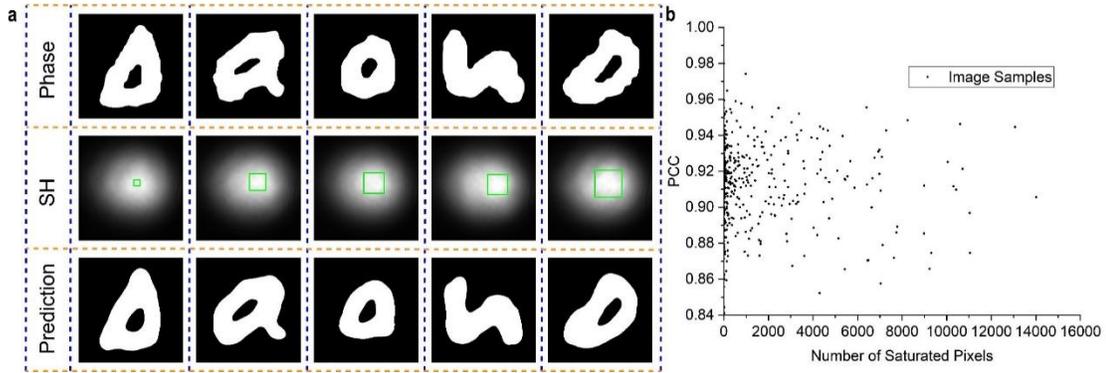

**Fig. S6. SHIELD performance with partially saturated input images.** (a) Test results for different overexposure. The green box shows the exposure area. (b) Quantified reconstruction fidelity with images containing different number of saturated pixels. No significant linear correlation between number of saturated pixels and the reconstruction fidelity in the test range (tested by linear regression assuming data points satisfied function y = ax + b, we get significance level p = 0.521 > 0.05 -- reject original assumption.).

## 5. Experimental layout

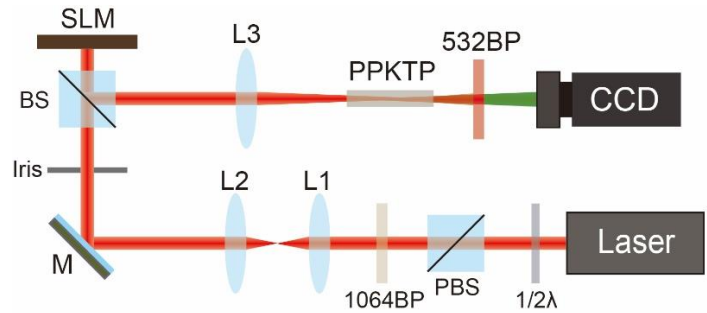

**Fig. S7. Experimental setup of SHIELD.** M and L represent mirrors and lenses, respectively. PPKTP: periodically poled potassium titanyl phosphate. PBS: polarizing beam splitter. BS: Beam Splitter.

## 6. Explanation of technical terms used

1.   Jaccard Index

JI (or the Jaccard similarity coefficient) is a statistic used in measuring the similarities between two sets, which is widely used in computer vision when measuring overlap in image recovery and segmentation. If two sets are identical, then the calculated Jaccard index equals to 1. On the contrary, if there is no overlap, the value equals to 0.

2.   Activation Function

A function that decides whether "activate" an artificial neuron or not based on its output value. It serves the purpose of introducing non-linearity into the output of a neuron. The implementation and the selection of activation function is critical for artificial neuron network to perform complex tasks.

3.   Loss Function

An objective function for comparing candidate solutions, and its value is minimized by neural network in the training process.

4. Concatenate

Concatenation is an operation to a list of tensors along dimension axis.

5. Depth to Space

An operation that generates a copy of the input tensor where values from the depth dimension are moved in spatial blocks to the height and width dimensions.

6. Epoch

One epoch is when an entire dataset is passed both forward and backward through the neural network. Each dataset will pass the neural network multiple times in the training process.

7. Underfitting and Overfitting

The underfitting when the neural network cannot adequately capture the underlying structure of the data, thus the fidelity cannot meet satisfaction. Similarly, the overfitting when the neural network demonstrates higher fidelity for known data in the learning process, but less accurate in predicting new data.

8. Batch Normalization

A method to get rid of internal covariate shift problem by normalizing the input of layers.

9. Learning Rate

A parameter that control how much adjustment the parameter each time for every step in training process.